\documentstyle[12pt,epsfig,epsf]{article}
\begin{document}

\vspace*{4cm}
{\baselineskip=15pt
\begin{center}
{\Large \bf LEPTOGENESIS WITH COSMIC STRINGS}
\end{center}
\par
\par
\begin{center}
{\large Rachel Jeannerot} 
       
{\em Centre for Theoretical Physics, University of Sussex,
Falmer, Brighton, BN1 9QH, UK. }
\end{center}
}
\vspace{1.5cm}
\begin{center}

{\em Published in the Proceedings of the XXXIInd Rencontres de Moriond
on  ``Electroweak Interactions and Unified Theories'', Les Arcs, Savoie, France,
March 15-22, 1997.}

\vspace{1.35cm}\end{center} 

\begin{center}
{\bf Abstract}
\end{center}

{\baselineskip=15pt
\noindent

A new scenario for baryogenesis is described. The basic idea is that theories 
beyond the standard 
model which contain a ${\rm U}(1)_{B-L}$ gauge symmetry, where $B$ and $L$ are 
respectively baryon and lepton numbers, 
predict the existence of {\em $B-L$ cosmic strings} with 
right-handed neutrinos trapped as transverse zero modes.
Cosmic string loops loose their energy via gravitational radiation and
rapidly decay releasing these
right-handed neutrinos. These decay into lepton and electroweak Higgs
boson producing an initial lepton asymmetry. This lepton asymmetry is then converted 
into a baryon asymmetry via sphaleron transitions. The minimal extension 
of the standard model and the minimal grand unified theory for which this scenario works 
are respectively
${\rm SU}(3)_c \times {\rm SU}(2)_L \times {\rm U}(1)_R \times {\rm U}(1)_{B-L}$ and SO(10).

}

\newpage

The aim of baryogenesis is to predict the matter/anti-matter asymmetry 
of the Universe. This is characterised by a parameter $\eta$ which is equal 
to the ratio of the baryon number density to the the photon number density 
${n_B \over n_\gamma}$
and is predicted by nucleosynthesis to be$^{1)}$
\begin{equation}
\eta = {n_B \over n_\gamma} = (2 - 7) \times 10^{-10} .
\end{equation}
Baryogenesis scenarios have to deal with the problems of 
sphalerons$^{2)}$. Sphaleron transitions violate $B+L$ and conserve 
$B-L$ ($B$ and $L$ are
respectively baryon and lepton numbers) and are very rapid between
$10^2$ and $10^{12}$ GeV$^{3)}$. Since 
$B = {(B+L)\over 2} + {(B-L)\over 2}$ and $L = {(B+L)\over 2} - 
{(B-L)\over 2}$we have:
$$
<B>_T \simeq  \beta \: <(B-L)>_T \hspace{0.5cm} {\rm and} \hspace{0.5cm}
<L>_T \simeq \gamma \: <(B-L)>_T 
$$
with $\beta$ and $\gamma$ close to $0.5$$^{4)}$. Hence we see that
unless the Universe started with a non-vanishing $B-L$ asymmetry, any baryon 
or lepton asymmetry generated at the grand unfied scale
will be erased by sphaleron 
transitions. 

In 1986, Fukugita and Yanagida$^{5)}$ introduced then the idea of 
leptogenesis: a lepton asymmetry is first produced 
which is then converted into a baryon 
asymmetry via sphaleron transitions. We present here a new scenario
for leptogenesis$^{6)}$.

The basic idea is the following. Theories beyond the standard model 
which contain a ${\rm U}(1)_{B-L}$
gauge symmetry, predict the existence of $B-L$ cosmic strings, that 
are cosmic strings associated with the breaking of 
${\rm U}(1)_{B-L}$,
as well as the existence of right-handed neutrinos
acquiring a Majorana mass at the $B-L$ breaking scale. As a consequence, there are
right-handed neutrinos trapped as transverse zero modes in $B-L$ cosmic string cores.
Cosmic string loops loose their energy emitting gravitationnal radiation
and rapidly decay releasing these neutrinos. These neutrinos
decay producing an initial 
lepton asymmetry.

Cosmic strings are one dimensional topological defects which form
when a gauge group 
G spontaneously breaks down to a subgroup
H of G if the vacuum manifold $G\over H$ contains non 
contractible loops, i.e. if the first homotopy group 
$\pi_1({G\over H}) $ is non trivial. They form, for example,
 when a {\rm U}(1) symmetry breaks down to the identity.
When a gauge group 
${\rm G} \supset {\rm U}(1)_{B-L}$
breaks down to a subgroup ${\rm H} \not\supset {\rm U}(1)_{B-L}$ of G,
$B-L$ cosmic strings form. 
The simplest
theory beyond the standard model which predicts the existence of
$B-L$ cosmic strings ${\rm SU}(3)_c \times  
{\rm SU}(2)_L \times {\rm U}(1)_R \times {\rm U}(1)_{B-L}$ and the
simplest grand unfied theory is SO(10). 

Now the Higgs field which forms a $B-L$ string, which we call $\Phi_{B-L}$,
is the same Higgs field which is used to give 
a superheavy Majorana mass to the right-handed neutrino. Therefore, 
there are right-handed neutrinos trapped 
as transverse zero modes$^{7)}$ in $B-L$ cosmic string core$^{6)}$. The zero mode
solution can be found by solving the equations of motion for the right-handed
neutrino field:
\begin{equation}
i \gamma^\mu D_\mu \nu_L^c - i \lambda \Phi_{B-L}^* \overline{\nu}_l^c = 0 
.
\end{equation}
$\overline{\nu}_l^c = C \gamma_0^T \nu_R^*$ 
transforms as a singlet under the standard model gauge group.
 $C$ is the charge conjugation matrix and $\lambda$ 
is a Yukawa coupling constant. 
The spinor $N = \nu_R + \nu_L^c$ is a Majorana spinor. For a straight 
infinite
cosmic string lying along the $z$-axis, 
$\Phi_{B-L} = f(r) e^{i \theta}$, with $f(0) = 0$ and 
$ f \stackrel{r \rightarrow \infty} \rightarrow
M_{B-L}$,
where $M_{B-L}$ is the $B-L$ breaking scale. The zero mode solution is then
given by:
\begin{equation}
N = \beta(r,\theta) \: \alpha(z+t) \label{eq:nuR}
\end{equation}
where $\beta(r,\theta)$ is a function peaked at $r=0$ which
exponentially vanishes outside the core of the string,
so that the fermions effectively live on the string. $\alpha(z+t)$ which
shows that the neutrinos travel at
the speed of light in the $-z$ direction, so that they are effectively
massless. 

On a straight string the energy to momentum relation $E = P$ holds, we have 
a continuous spectrum of states and the Fermi energy $E_F = 0$. However,
on a cosmic string loop of radius $R$, the energy relation 
must be modified to$^{8)}$: $E = {(L+{1\over 2})\over R} = P + {1\over 2R}$,
where $L$ is the neutrino angular momentum. The neutrino energy spectrum is 
then $E ={ \pm ({\rm n}  + {1\over 2})  \over R} $
where $n$ is an integer, we have a discrete spectrum of states and the
Fermi energy $E_F = {1\over 2 R} $.

When a string network forms, some closed loops are initially formed 
as well as 
infinite strings. Cosmic string loops loose their energy via gravitational 
radiation at a rate ${\dot{E}} = - \Gamma_{\rm loops} {M_{B-L} \over M_{pl}}$
and rapidly decay. The network evolves, and more
loops form via the intercommuting of long strings. 
A loop decays when its radius becomes 
comparable to its width $\sim 
M_{B-L}^{-1}$. The neutrino Fermi energy is then 
$E_F \sim {1\over 2} M_{B-L} $, which 
is lower than the energy needed by
a neutrino to escape the string. Hence, when a cosmic string loop decays, it 
releases at least $n_\nu = 1$ heavy Majorana neutrino. This is an out-of-equilibrium 
process.   

The released neutrinos $N$ decay according to the two diagrams
shown in figure 1, producing a lepton asymmetry.
\begin{figure}
\centerline{\psfig{file=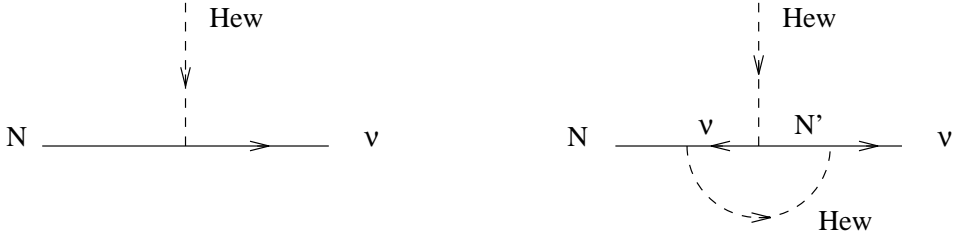}}
{\small \caption{The tree level and one loop diagrams for
right-handed neutrino decay.}}
\end{figure}
CP is violated through the one loop radiative
correction involving an electroweak Higgs 
particle. The
lepton asymmetry produced is characterised by the CP violation parameter$^{9)}$:  
\begin{equation}
\epsilon \simeq {m_{D_3}^2 \over \pi v^2} {M_{N1} \over M_{N2}} \sin{\delta
  } \, ,
\end{equation}
where $m_{D_3}$ is the Dirac mass of the third lepton generation, $v$
is the vacuum expectation value of the electroweak Higgs field, $v = 
<H_{ew}> = 174$ GeV, $M_{N1}$ and $M_{N2}$ are the right-handed
neutrino Majorana masses of the first and second generation respectively and
$\delta $ is the CP violating phase. 

The generated baryon number  
 per comoving volume at temperature $T$ is then given by 
\begin{equation}
B(T) =  {1\over 2} {N_{\nu}(t) \epsilon \over s} , \label{eq:B-L}
\end{equation}
where s is the entropy of the Universe at time $t$ and 
$N_\nu(t)$ is  $n_\nu$ times the number density of cosmic string  
loops which have shrinked to a point at time t.

The final baryon asymmetry is therefore a function of the cosmic string scenario
parameters, of the neutrino mass matrix and of the strength of CP violation.
Using the model of Copeland et al.$^{10)}$ for cosmic string loop evolution,
and assuming that the Dirac neutrino masses fall into a hierarchical pattern 
similar to that of leptons and quarks, we find:
\begin{eqnarray}
B_{\rm final} \simeq {13.8 \over ( 0.3 \: \pi)^3} \: {\nu_s \over (k-1)
k^{3\over 2 } \Gamma_{loops}} {M_{B-L}
  \over M_{{\rm pl}}} \: {m^2_{D_3} \over v^2}\: {M_{N1} \over M_{N2}} \:
\sin{\delta } 
\end{eqnarray}
where $\nu_s$  and $k$ are respectively related to the different length 
scales in the string network 
 and to the life-time of a loop.

Fixing 
${M_{N1} \over M_{N2}} = 0.1$, assuming $m_{D_3} =1 - 100$
GeV and assuming maximum CP violation,
i.e. $\sin { \delta } = 1$, we find that the parameter $\eta$  
is predicted with the $B-L$ breaking scale in the range 
\begin{equation}
M_{B-L} =  (1 \times 10^{6} - 2 \times 10^{15})  \; {\rm GeV} . 
\end{equation}
The minimal extension of the standard model for which this scenario works
is ${\rm SU}(3)_c \times {\rm SU}(2)_L \times {\rm U}(1)_R \times {\rm U}(1)_{B-L}$ and 
the minimal GUT is SO(10).

\vspace{.5cm}

I would like to thank the Conference organisers and everybody who made sure 
that the conference went smoothly. I would also like to thank M. Shaposhnikov
and G. Senjanovic for discussions.

\vspace{.5cm}

{\baselineskip=15pt

\noindent 1.~C.J. Copi, D.N. Schramm and M.S. Turner, Science 267, 192 
(1995); {\em ibid.}, Phys. Rev. Lett. 75, 3981 (1995).

\noindent 2.~N. Manton. Phys. Rev. D28, 2019 (1983);
  F. Klinkhamer and N. Manton, Phys. Rev. D30, 2212 (1984).  
  
\noindent 3.~V.A. Kuzmin, V.A. Rubakov and M.E. Shaposhnikov,
  Phys. Lett. B155, 36 (1985).  
  
\noindent 4.~S.Yu. Khlebinov and M.E. Shaposhnikov, Phys. Lett. B387, 817 (1996).

\noindent 5.~M. Fukugita and T. Yanagida, Phys. Lett. B174, 45
  (1986). 
  
\noindent 6.~R. Jeannerot, Phys. Rev. Lett. 77, 3292 (1996).

\noindent 7.~R. Jackiw and P. Rossi, Nucl. Phys. B190, 681
  (1981). 
  
\noindent 8.~S.M. Barr and A.M. Matheson, Phys. Lett.
B198, 146 (1987).

\noindent 9.~M.A. Luty, Phys. Rev. D45, 455 (1992).

\noindent 10.~D. Austin, E.J. Copeland and T.W.B. Kibble,
  Phys. Rev. D51, 2499 (1993).

}
\end{document}